# Harmonic content analysis of a soft starting variable frequency motor drive based on FPGA

Yogesh Sapkota
*Electrical and Computer Engineering*
*Youngstown State University*
Youngstown, OH, USA
ysapkota@student.ysu.edu

Suman Devkota
*Electrical and Computer Engineering*
*Youngstown State University*
Youngstown, OH, USA
sdevkota01@student.ysu.edu

Vamsi Borra
*Electrical and Computer Engineering*
*Youngstown State University*
Youngstown, OH, USA
vsborra@ysu.edu

Pedro Cortes
*Chemical Engineering*
*Youngstown State University*
Youngstown, OH, USA
pcortes@ysu.edu

Srikanth Itapu
*Department of ECE*
*Alliance University*
Bengaluru, India
srikanth.itapu@alliance.edu.in

Frank Li
*Electrical and Computer Engineering*
*Youngstown State University*
Youngstown, OH, USA
xli@ysu.edu

*Abstract*—As the demands for electric vehicles, electric aircrafts, unmanned aircraft systems, and other motor-driven systems increase, high-performance motor drives employing variable frequency control with higher efficiency a nd r eliability are becoming increasingly important parts of the ever-changing technological landscape. This study proposes a Field Programmable Gate Array (FPGA)-based variable frequency soft-starting motor drive for a three-phase induction motor. Variable frequency drive is realized by employing Direct Digital Synthesis (DDS) technique. Gradually increasing the modulation index during the startup enables the soft start of the induction motor to limit the inrush current and controlling the soft-start period enables the reduction of the peak in-rush current. Digilent Arty A7-35T Artix-7 FPGA Development board is used to realize the proposed motor drive. The performance of three different Pulse Width Modulation (PWM) techniques—Sinusoidal PWM, Third Harmonic Injected PWM, and Space Vector PWM are analyzed and compared. The proposed architectures require only a small fraction of the FPGA's resources. The inverter output voltage and the load currents are analyzed for the harmonic contents using MATLAB. In the experimental realization, a four-pole squirrel cage delta- connected induction motor is utilized with a switching frequency of 4 kHz. The current and voltage characteristics of the induction motor are studied under different operating conditions to study harmonic contents and the effect of changing soft-start duration. The findings demonstrate a low-cost, flexible control of the induction motor with improved harmonic performance.

## I. Introduction

Field Programmable Gate Arrays' (FPGA) usage in power electronics have been explored since the late 1990s [1]. As the demand for higher frequency switching, along with lower conduction losses grew, FPGA's highly parallel processing prowess attracted the interest of researchers and power electronic engineers. FPGA's reconfigurable nature along with higher throughput, in addition to the moderate power consumption, has made it a lucrative alternative to ASICs and microcontrollers for prototyping and testing new algorithms in the field of power electronics [2]. Likewise, multiple cores of FPGA can be used to control multiple motor control systems parallelly. Three-phase induction motors are ubiquitous in industrial application because of their simple, low cost and rugged construction [3]. Self-starting capabilities and absence of brushes like in brushed DC motor lends to easier maintenance and longevity of the induction motors. Despite these benefits, induction motors have difficult control schemes for speed control. Advancements in the fields of motor drives with self-starters and variable frequency drives have eased the speed control for three phase induction motors. FPGA's high computational prowess can be harnessed for better and precise speed control of the three-phase induction motor.

This research paper is structured as follows: Section II briefly discusses about the PWM techniques implemented in this study, Section III elaborates on the FPGA architecture of the variable frequency PWM generators, and Section IV presents the findings from the experimental observations.

## II. Pulse Width Modulations (PWM) Techniques

In comparison to the conventional Voltage Source Inverter (VSI) like square wave inverters, which only allow the frequency control of the output voltage, PWM techniques offer versatile control of the amplitude and frequency of the output voltage along with control of harmonic content present in the output voltage [4]. PWM techniques are preferred for the DC-AC converters especially in three-phase applications because they can reduce the harmonics and switching losses [2]. Sinusoidal PWM (SPWM), Third Harmonic Injection (THISPWM) and Space-Vector PWM (SVPWM) are commonly employed PWM inverter techniques. These PWM techniques can be modified depending upon the requirement of the three-phase loads to focus on different areas of power conversion.

Navarro et. al. [5] have focused on high resolution output from SPWM. PWM modified to achieve high switching frequency was discussed by Lakka et. al [6]. SPWM has been modified to PD-SPWM [7] with a goal to reduce harmonic content and increase output voltage. Sarker et. al propose a high definition SPWM (HD-SPWM) technique that aimed at suppressing inverter output harmonics [8]. Carrier based PWM, where a carrier waveform of high frequency is compared against a reference waveform, is a widely implemented technique. Sinusoidal PWM, THI-SPWM along with carrier-based equivalence of Space Vector PWM are briefly discussed in this section.

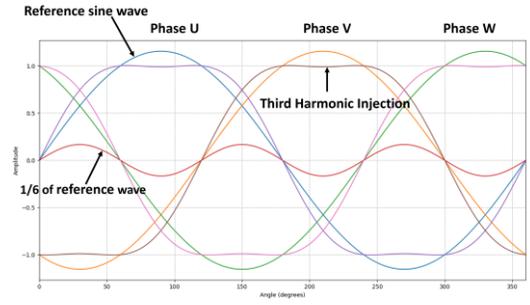

Fig. 2. Modification of SPWM to THI-SPWM by addition of third harmonic content.

### A. Sinusoidal PWM (SPWM)

In the SPWM technique, a sinusoidal waveform, called the reference signal $v_r$, of the desired output frequency, $f_r$, is compared against a carrier waveform $v_c$ with a switching frequency, $f_s$, to determine the switching states of the gates driving the inverter output. The output voltage delivered to the motor phases is determined by modulation index (m) which is given as:

$$m = v_r / v_c \qquad (1)$$

and, the peak value of the magnitude of the output voltage is given as,

$$v_o = m \cdot V_{dc} \qquad (2)$$

where, $V_{dc}$ is the DC link voltage.

### C. Space Vector PWM (SVPWM)

Implementation of conventional Space Vector PWM requires a complex algorithm; however, the traditional SPWM method can be modified to generate gating pulses identical to the those produced by conventional SVPWM [10]. The modified modulating waveform is obtained by adding a common mode voltage which is given by,

$$v_{cmv} = -\frac{(v_{max} + v_{min})}{2} \qquad (4)$$

$v_{max}$ and $v_{min}$ are the maximum and minimum values of the three phases of the sine ($v_a$, $v_b$, $v_c$) at any given angle and are given as:

$$v_{max} = max(v_a, v_b, v_c), v_{min} = min(v_a, v_b, v_c) \qquad (5)$$

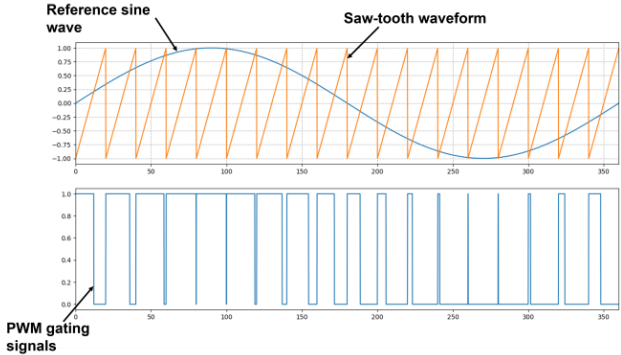

Fig. 1. Comparison of sine wave and saw-tooth wave to produce switching signals.

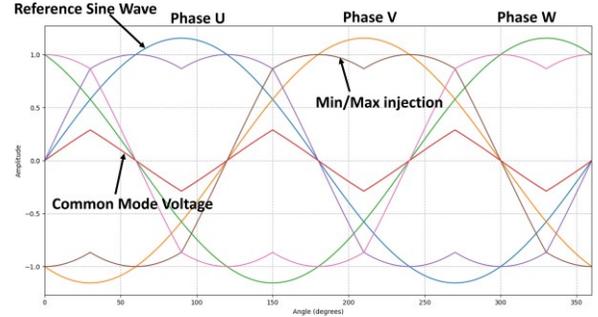

Fig. 3. Modification of SPWM to SVPWM by addition of common mode voltage.

### B. Third Harmonic Injected Sinusoidal PWM (THI-SPWM)

Conventional SPWM can be modified according to (3) by adding a third harmonic component to obtain the modulating signal for THI-SPWM [9] as shown in Fig. 2.

$$y = 1.155 sin(\omega t) + \frac{1}{6} sin(3\omega t) \qquad (3)$$

Third harmonic injection using (3) results in better utilization of the DC bus along with reduced harmonic content in the output voltage [9].

SVPWM reduces the harmonic contents present in the voltage and current output waveform [11].

### III. FPGA ARCHITECTURE

Fig. 4 depicts a simplified block diagram of the proposed SPWM generator architecture. It consists of a 100 MHz clock, a sine Look-Up table of size 3600 elements, and saw-tooth waveform implemented using an up-counter, and a comparator. In addition, it has a variable frequency output block that generates 5-100 Hz sine output and a variable modulation index based soft-starting block that controls the soft-startup duration.

The values for sine LUT are generated using MATLAB and Python. A finite state machine is developed to realize the SPWM driver. A summary of the functionality of the proposed SPWM generator is illustrated in Fig. 5.

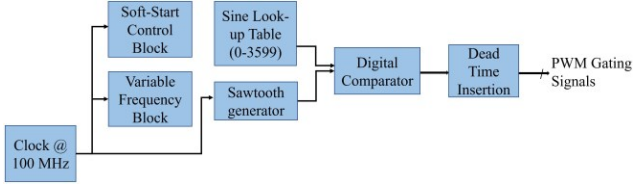

Fig. 4. Simplified block diagram of the proposed architecture.

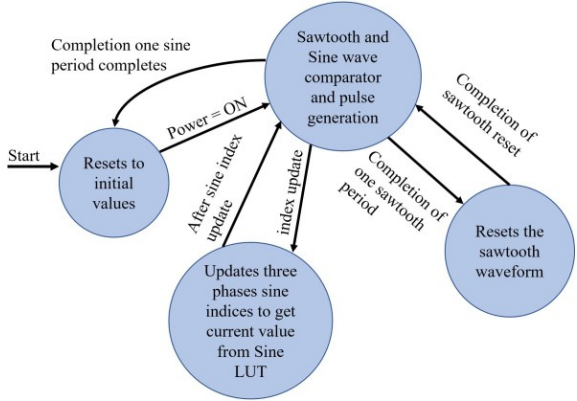

Fig. 5. SPWM generator workflow.

In this setup, soft startup is realized by gradually increasing the modulation index. The sine look up table offers $0.1°$ resolution in the sine value, with values ranging from 125000 to 150000 and a zero-crossing point at 137500. The carrier wave has a frequency of 4 kHz. Variable frequency is achieved by drawing from the concepts of direct digital synthesizer (DDS). DDS is a technique to realize an analog signal using digital system. A typical DDS system consists of a reference clock, a phase accumulator, phase to amplitude conversion, usually a look-up table, along with a Digital to Analog converter and a low pass filter [12]. Sine waves with micro-Hz resolution [13] and a large range of output frequency [14] can be synthesized with DDS. In this proposed architecture, an equivalent model of DDS is proposed, where the phase increment is implemented using a sine index counter whose value depends on the frequency similar to the phase increment of DDS. The proposed architecture only uses a fraction of the FPGA's resources as shown in Table (I).

## IV. EXPERIMENTAL SETUP AND RESULTS

In the experimental realization, a four-pole squirrel cage delta-connected induction motor is utilized. The reference signal is a 60 Hz sine wave. The VHDL code was developed using Xilinx Vivado and the generated bitstream was subsequently downloaded to Digilent Arty A7 FPGA Development board.

TABLE I
FPGA RESOURCE UTILIZATION

| Resource | Utilization | Available | Utilization % |
|---|---|---|---|
| LUT | 385 | 20800 | 1.85 |
| FF | 364 | 41600 | 0.88 |
| BRAM | 4 | 50 | 8.00 |
| IO | 30 | 210 | 14.29 |
| BUFG | 2 | 32 | 6.25 |

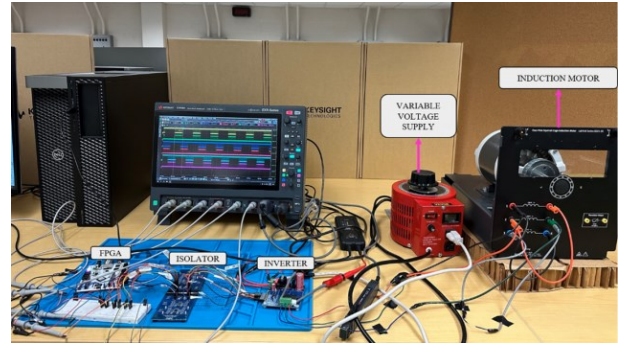

Fig. 6. Experimental setup.

In Fig. 6, a variable voltage supply was used to provide a 120 V DC bus supply to the inverter which was connected to three phases of the 200 W delta-connected induction motor. The motor is controlled using an open-loop control scheme. An isolator is connected between Infineon's Eval-M1-05-65D inverter evaluation board and the FPGA to separate the low-voltage and high-voltage side. The 4 kHz switching frequency gating pulses shown in Fig. 7 were captured using oscilloscope. The motor line-line voltage waveform along with the line and phase current waveforms were collected and analyzed using MATLAB/Simulink.

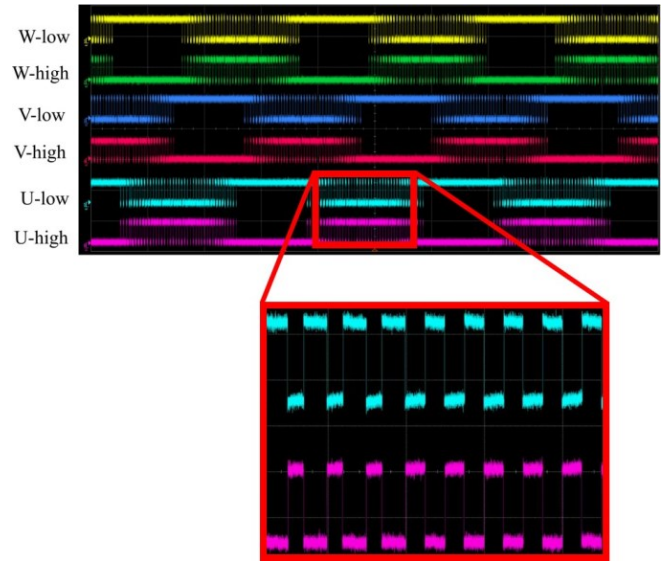

Fig. 7. U,V and W phases SPWM gating signals for $f_{sw}$ = 4 kHz, $f_r$ = 60 Hz and m = 0.6.

## A. Harmonic Content Analysis

The harmonic content present in the motor phase voltage along with motor line and phase current are studied using Fast Fourier transform (FFT) and Total harmonic distortion (THD) functions of MATLAB. The phase current and phase voltage THD values for m = 0.4 are shown in Fig. 8. It is observed that the THD is lower for THISPWM and SVPWM compared to conventional SPWM. Modification of the modulating signal provides a simple and effective method of improving the harmonic contents of the output voltage and current waveform.

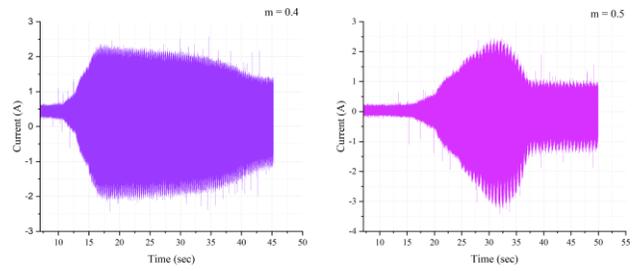

Fig. 9. Startup current for m = 0.4 and 0.5.

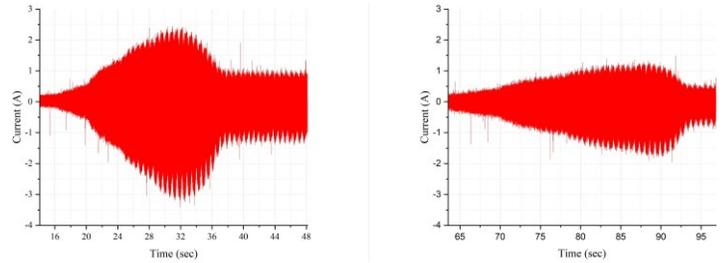

Fig. 10. Controlling the amplitude of peak startup currrent by changing the duration of soft-start.

## V. Conclusion

A simple to implement, low cost and flexible FPGA based variable frequency motor drive for three-phase induction motor capable soft-starting with three different PWM methods- SPWM, THI-SPWM, SVPWM-is proposed. The proposed architecture utilizes a small fraction of the FPGA's resources. Analysis of the experimental results showed improved performance in terms of harmonic content with reduced THD for THI-SPWM and SVPWM compared to conventional SPWM. A simple modification in the FPGA's architecture allowed us to switch between these PWM techniques. In addition, the peak current control of at the startup was also achieved. This proposed method can be implemented in a wide range of induction motors with higher power outputs and improved performance.

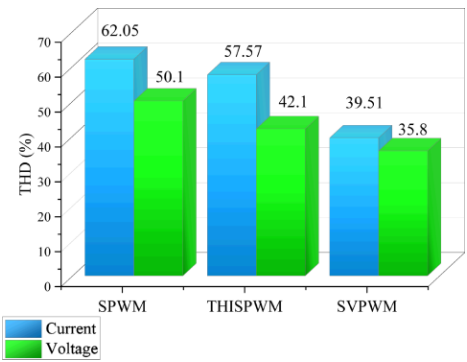

Fig. 8. Motor phase current and phase voltage THD comparison of SPWM with THISPWM and SVPWM $f_{sw}$ = 4 kHz, $f_r$ = 60 Hz and m = 0.4.

## B. Soft start up analysis

It is noted that soft-start up time varies as the modulation index changes, which can be observed in Fig. 9. The motor phase voltage increases as the modulation index increases. As a result, the motor reaches nominal no-load speed at different times with different motor no-load current for different values of modulation indices. Fig. 10 shows that by adjusting the soft-startup period, the current delivered to the motor terminals can be controlled. The in-rush current supplied to the motor phases can be controlled by gradually increasing the pulse-width at different rates .The soft-start control design reduces the startup peak-peak (pk-pk) current as shown in Table (II).

### Acknowledgment

This work has been carried out with the support from the U. S Air Force, via Assured Digital Microelectronics Education and Training Ecosystem (ADMETE) grant (FA8650-20-2-1136).

TABLE II
Peak-Peak Current Comparison

| Reference | Startup pk-pk current (A) | Nominal pk-pk current (A) |
|---|---|---|
| 1 | 2.98 | 1.19 |
| 2 | 5.35 | 2.28 |

### References

[1] A. Al-Safi, A. Al-Khayyat, A. M. Manati, and L. Alhafadhi, "Advances in FPGA Based PWM Generation for Power Electronics Applications: Literature Review," in 11th Annual IEEE Information Technology, Electronics and Mobile Communication Conference, IEMCON 2020, Nov. 2020, pp. 252–259. doi: 10.1109/IEMCON51383.2020.9284821
[2] Y.-Y. Tzou and H.-J. Hsu, "FPGA Realization of Space-Vector PWM Control IC for Three-Phase PWM Inverters," 1997. [Online]. Available: http://pemclab.cn.nctu.edu.tw
[3] N. Mohan and S. Raju, Power Electronics, A First Course: Simulations and Laboratory Implementations. John Wiley & Sons, 2022.

[4] S.-H. Kim, "Pulse width modulation inverters," in Electric Motor Control, Elsevier, 2017, pp. 265–340. doi: 10.1016/b978-0-12-812138-2.00007-6.

[5] D. Navarro, Ó. Lucía, L. A. Barragán, J. I. Artigas, I. Urriza, and Ó. Jiménez, "Synchronous FPGA-based high-resolution implementations of digital pulse-width modulators," IEEE Trans Power Electron, vol. 27, no. 5, pp. 2515–2525, 2012, doi: 10.1109/TPEL.2011.2173702.

[6] M. Lakka, E. Koutroulis, and A. Dollas, "Development of an FPGA-based SPWM generator for high switching frequency DC/AC inverters," IEEE Trans Power Electron, vol. 29, no. 1, pp. 356–365, 2014, doi: 10.1109/TPEL.2013.2253216.

[7] X. Xianglian, X. Pingting, T. Gang, T. Zilin, and D. Diankuan, FPGA based multiplex PWM generator for diode-clamped cascaded inverter in the direct-driven wind power system.

[8] R. Sarker, A. Datta, and S. Debnath, "FPGA-Based High-Definition SPWM Generation with Harmonic Mitigation Property for Voltage Source Inverter Applications," IEEE Trans Industr Inform, vol. 17, no. 2, pp. 1352–1362, Feb. 2021, doi: 10.1109/TII.2020.2983844.

[9] J. A. Houldsworth and D. A. Grant, "The Use of Harmonic Distortion to Increase the Output Voltage of a Three-Phase PWM Inverter," IEEE Trans Ind Appl, vol. IA-20, no. 5, pp. 1224–1228, 1984, doi: 10.1109/TIA.1984.4504587.

[10] P. S. Varma and G. Narayanan, "Space vector PWM as a modified form of sine-triangle PWM for simple analog or digital implementation," IETE J Res, vol. 52, no. 6, pp. 435–449, 2006, doi: 10.1080/03772063.2006.11416484.

[11] K. V. Kumar, P. A. Michael, J. P. John, and S. S. Kumar, "Simulation and comparison of SPWM and SVPWM control for three phase inverter," ARPN journal of engineering and applied sciences, vol. 5, no. 7, pp. 61–74, 2010.

[12] H. Omran, K. Sharaf, and M. Ibrahim, "An all-digital direct digital synthesizer fully implemented on FPGA," in 2009 4th International Design and Test Workshop, IDT 2009, 2009. doi: 10.1109/IDT.2009.5404133.

[13] Analog Device, Inc., "A Technical Tutorial on Digital Signal Synthesis," 1999.

[14] W. Pietrowski, W. Ludowicz, and R. M. Wojciechowski, "The wide range of output frequency regulation method for the inverter using the combination of PWM and DDS," COMPEL - The International Journal for Computation and Mathematics in Electrical and Electronic Engineering, vol. 38, no. 4, pp. 1323–1333, Aug. 2019, doi: 10.1108/COMPEL-10-2018-0402.